# Near 13% efficient semitransparent Cu(In,Ga)S$_2$ solar cells with band gap of 1.6 eV on transparent back contact

Kulwinder Kaur[1]*, Arivazhagan Valluvar Oli[1], Michele Melchiorre[1], Wolfram Hempel[2], Wolfram Witte[2], Jan Keller[3], Susanne Siebentritt[1]

[1]Laboratory for Photovoltaics, Department of Physics and Materials Science, University of Luxembourg, Luxembourg

[2]Zentrum für Sonnenenergie- und Wasserstoff-Forschung Baden-Württemberg (ZSW), Germany

[3]Ångström Solar Center, Division of Solar Cell Technology, Uppsala University, Sweden

*Corresponding author: kulwinder.kaur@uni.lu

**Abstract**

Wide-gap Cu(In,Ga)S$_2$ solar cells with In$_2$O$_3$:Sn (ITO) as transparent back contact are evaluated for the application as top cells in tandem devices. The effect of Na on the solar cell performance is investigated by supplying additional Na by NaF co-evaporation or exclusively by Na diffusion from glass. An efficiency of 12.7% is achieved for a semitransparent solar cell with a band gap of 1.6 eV, with sufficient Na diffusion from glass only, allowed by a thin ITO layer. Absorber grown with additional NaF co-evaporation during Cu(In,Ga)S$_2$ growth on thicker ITO show a comparable efficiency of 12%. High temperature growth at $T_{sub}$ = 630°C enhances overall absorber quality and results in wide-gap absorbers, with photoluminescence quantum yield improved to 1.5×10$^{-5}$, two orders of magnitude higher than absorber grown at low temperature. NaF co-evaporation is effective in suppressing deep defects, thereby reducing non-radiative recombination and enhancing photoluminescence quantum yield further. A GaO$_x$ interfacial layer is formed at the rear contact, likely contributing to the passivation of the back contact. With the presence of thick GaO$_x$ layer, current blocking effects are visible in the current-voltage curves. On the contrary, a thinner ITO tends to result in thinner GaO$_x$ layer and no current blocking is observed.

## 1. Introduction

Chalcopyrite solar cell technology boasts a record efficiency of 23.6% for $(Ag,Cu)(In,Ga)Se_2$ (ACIGSe) solar cells with band gap of 1.1 eV [1]. Wide-gap sulfide chalcopyrite $Cu(In,Ga)S_2$ (CIGS) solar cells with band gaps around 1.5 eV or above have also gained attention because of their potential application as top cell in tandem devices [2-4]. For the top cells, NIR transparency of the rear contact is required to transmit light to lower band gap bottom cell (typically Si bottom cell), therefore a transparent back contact (TBC) is needed instead of the conventional opaque Mo back contact. There are reports exploring the potential of ITO ($In_2O_3$:Sn) [5-9], IOW ($In_2O_3$:W) [10], and IOH ($In_2O_3$:H) [8, 11, 12], as possible TBCs for CIGS top cells. Recently, wide band gap ACIGSe with high silver content AAC ([Ag]/[Ag]+[Cu]) of 0.8 solar cells have shown efficiency of 13.6% with a band gap of 1.44 eV using IOW as back contact [10]. Both 2T and 4T tandem configurations have distinct benefits and drawbacks, however, selecting an appropriate bandgap combination is essential for tandem devices to exceed theoretical limits. To realize highest efficiencies with 1.1 eV bottom cell, a top cell with a band gap of 1.6 eV or higher is needed [13, 14]. Higher Ga content is required to make $Cu(In,Ga)Se_2$ (CIGSe) a wide-gap absorber that eventually reduces the absorber quality. It is reported that high Ga absorbers exhibit low diffusion lengths which is unfavorable for efficient carrier collection and ascends the $V_{OC}$ deficit [15-17]. Therefore, sulfide based wide-gap CIGS with tunable band gap from 1.5 eV to 2.4 eV is a viable choice for the application of top cells of tandem devices [18]. However, a high quasi-Fermi level splitting (QFLS) and high efficiency in CIGS single junction devices are prerequisites for the best usage of CIGS in tandem applications.

CIGS owns the chalcopyrite structure, same as CIGSe, still the certified record efficiency of CIGS so far is not higher than 15.5% with Mo back contact with a band gap of 1.55 eV [3]. Previous work from our lab demonstrated in-house efficiency of 16.1% (active area) and certified efficiency 14.8% with band gap of 1.6 eV by mitigating the composition segregation with Mo back contact [19]. It is vital to comprehend the causes of losses in sulfide CIGS absorbers to enhance their performance. One of the primary losses in chalcopyrite absorbers is non-radiative loss, which eventually restricts the ability to generate charge carriers effectively and reduces steady state carrier density. The latter has a direct relationship with the absorber's QFLS and lifetime of the charge carriers [20]. QFLS is a critical parameter for assessing absorber quality, as it defines the theoretical upper limit of the open-circuit voltage ($V_{OC}$) achievable from a solar cell utilizing that absorber. Non- radiative losses in CIGS absorbers are associated to both bulk and interfacial loss. Deep defects in the bulk of CIGS act as recombination centers [21]. Shockley-Read-Hall (SRH) recombination describes the recombination at such deep states and is especially detrimental to solar device performance. Lumoscio et al. showed that QFLS remains lower than S-Q limit of $V_{OC}$ in Cu-poor and Cu-rich $CuInS_2$ absorbers which is mainly limited due to deep defects. In addition, high deposition temperature can reduce these defects, which suppresses non-radiative recombination and enhances the absorber's QFLS [22]. However, high deposition temperature may degrade both electrical and optical properties of the TCOs used for the back contact. In addition, formation of gallium oxide $GaO_x$ has been observed between the CIGS and TBC interface which may act as an extraction barrier for the charge carriers [5, 11, 23]. Yang et al. have observed presence of $GaO_x$ interlayer at CIGSe/ITO interface even at growth temperature of 453°C [5]. On the contrary, using FTO as a back contact showed no formation of $GaO_x$ at temperature as high as 550°C [6, 24]. Also, it has been reported that the presence of Na and the thickness of ITO play an important role in the formation of $GaO_x$ [9, 25, 26]. Whether $GaO_x$ negatively impacts solar cell performance remains an open question, as various research groups have reported high efficiencies even with a $GaO_x$ interfacial layer at the back contact. [7, 8, 11]. Therefore, PCE appears to depend on the characteristics of the $GaO_x$ layer such as thickness, or type of doping in $GaO_x$. There is sufficient literature available on Se-based chalcopyrite solar cells with TBCs [5-12], however, reports on sulfide-based CIGS solar cells incorporating TBCs are comparatively scarce. Choubrac et. al. recently demonstrated efficiency of 11.1% under rear illumination with TBC and opaque top contact in CIGS solar cells for their potential application in photoelectrochemical cell systems [27].

In this work, ITO is investigated as TBC in sulfide chalcopyrite solar cells at high growth temperature of $T_{sub}$ ~630°C and compared to lower temperature $T_{sub}$ ~575°C processed solar cells. The role of sodium is also investigated by introducing NaF through co-evaporation or Na diffusion from the glass substrate. Significant improvement in photoluminescence (PL) quantum yield ($Y_{PL}$) is realized for absorbers grown at ~630°C compared to those grown at ~575°C. Introduction of Na effectively reduced the deep defects and lowers non-radiative recombination revealed by room temperature steady state PL studies. Thicker ITO resulted in thick $GaO_x$ at high temperature and Na at the interface has been observed to be beneficial for charge transport. Despite the presence of

GaO$_x$, PCE as high as 12.7% (active area) is demonstrated. The present work reflects the potential of sulfide-based chalcopyrite solar cells for the application of top cells in tandem architectures.

## 2. Results and Discussion

We examine sulfide chalcopyrite absorbers grown on ITO back contact of thickness 250 nm and 100 nm with sheet resistance of 14 Ω/sq and 30 Ω/sq respectively before absorber deposition. CIGS absorbers were grown by three stage co-evaporation [28], at two different substrate temperatures ($T_{sub}$=575°C and 630°C), following the recipes described in [19]. For further details we refer to experimental section. These substrate temperatures are referred to the maximum growth temperature during the CIGS growth process on ITO coated soda-lime glass (SLG). Na is supplied through the SLG into the CIGS absorber. In some samples additional Na is provided by NaF co-evaporation during the second stage. To control the amount of Na in the absorbers, the duration of NaF co-evaporation was controlled. We will compare samples T575/(0NaF)/ITO250, T630/(0Na)/ITO250, T630/(7Na)/ITO250, T630/(18Na)/ITO250, and T630/(0NaF)/ITO100, where the starting number represents the growth temperature, number in parentheses reflects NaF co-evaporation duration in minutes, on two different ITO thickness 250 nm and 100 nm, the schematic is depicted in Figure 1. All samples have a [Ga]/[Ga]+[In] ratio of about 25% and are slightly Cu-poor, besides the low temperature sample. The detailed chemical composition of the films is given in supplementary information, Table S1. All the CIGS absorbers are approximately 1.5 μm thick, as estimated from scanning electron microscope (SEM) cross-sectional images from different spots of the sample.

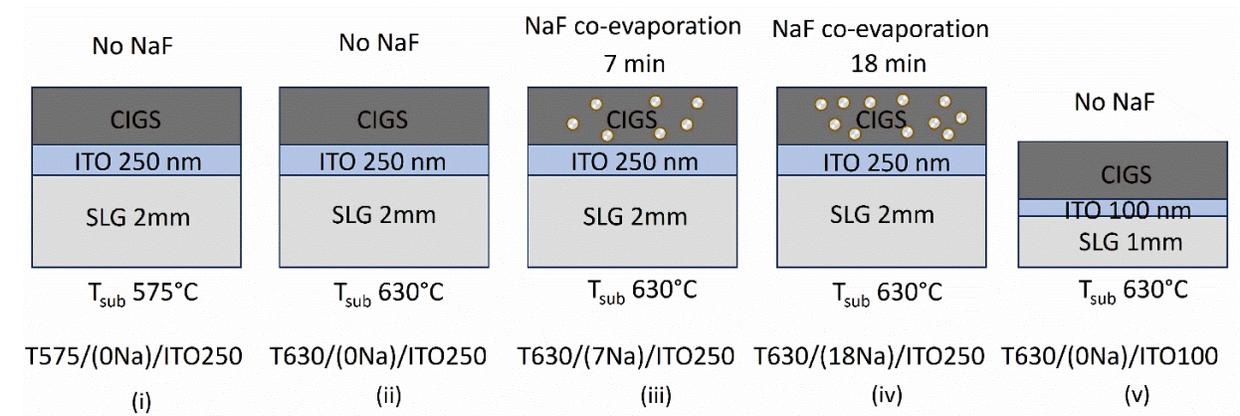

*Figure 1:* Schematic of the samples grown under different conditions which includes substrate temperature ($T_{sub}$), NaF co-evaporation and different ITO thicknesses, (i) T575/(0Na)/ITO250, (ii) T630/(0Na)/ITO250, (iii) T630/(7Na)/ITO250, (iv) T630/(18Na)/ITO250, (v) T630/(0Na)/ITO100.

## 2.1 Characterization of Cu(In,Ga)S$_2$ absorbers

### 2.1.1. Structural and compositional properties

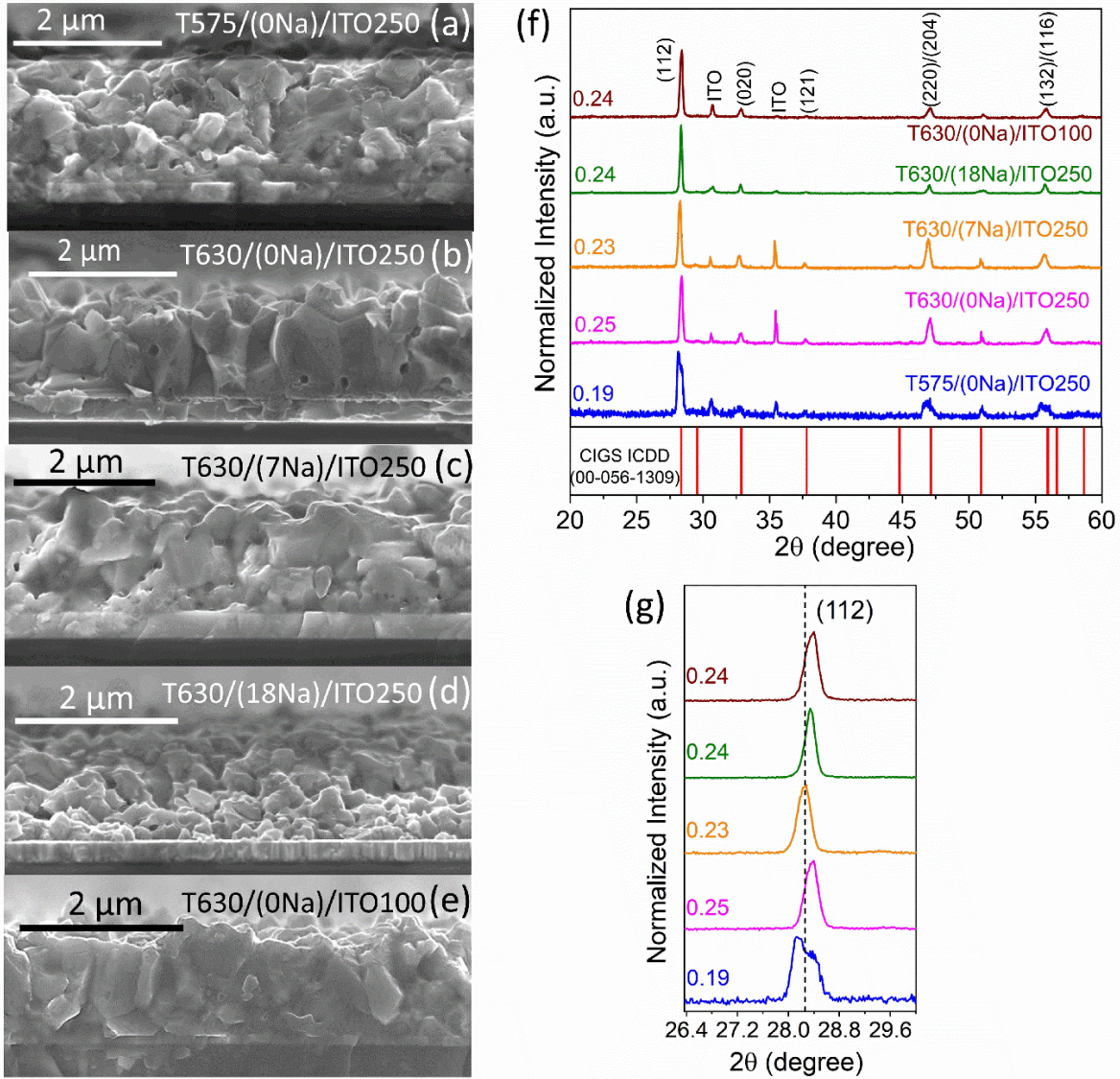

*Figure 2:* a) to e) SEM cross-sections of TBC/CIGS at different $T_{sub}$ with and without additional Na supply f) XRD diffraction pattern of CIGS absorbers grown at $T_{sub}$ ~575°C and ~630°C with different Na supply, respective [Ga]/([In]+ [Ga]) is marked along the spectra, g) enlarged view of (112) reflection. Dashed line represents the (112) peak position corresponds to ICDD (00-056-1309).

Figure 2(a-e) represent cross-section images from SEM of the CIGS absorbers grown at 575°C and 630°C with and without additional Na supply. When comparing absorbers grown under identical Na and ITO conditions, the substrate temperature has a pronounced impact on the microstructure. The absorber deposited at the lower substrate temperature (T575/(0Na)/ITO250) exhibits smaller grains, whereas increasing the substrate temperature to 630 °C (T630/(0Na)/ITO250) results in significantly larger, columnar grains with a reduced density of grain boundaries. This behavior is consistent with improved grain coalescence and vertical grain growth due to higher thermal energy available during growth process [29]. The apparent Cu-rich composition indicated by EDX for the absorber deposited at 575 °C is most likely influenced by the depth-averaged nature of the measurement, which can lead to an overestimation of the near-surface Cu content. In contrast, the observed microstructural features, specifically the relatively small grain size which are more consistent with Cu-poor growth conditions. This interpretation is supported by extensive reports on CuInS$_2$ absorbers, where Cu-poor compositions typically result in limited grain growth, while Cu-rich growth conditions promote significantly larger grains. Therefore, despite the EDX-derived composition, the microstructural evidence suggests that the 575 °C absorber is effectively Cu-poor during growth [2]. When comparing absorbers grown at the same substrate temperature, the Na incorporation level has a pronounced effect on the microstructure. In the present study, the absorber with moderate Na supply (7Na) exhibits larger grains, whereas the absorber with higher Na supply (18Na) shows significantly smaller grains and a higher density of grain boundaries compared to the other high-temperature absorbers. This trend suggests that while moderate Na incorporation promotes grain growth, excessive Na adversely affects grain coalescence. Sodium is often observed to enhance grain growth in CIGS thin films, particularly when starting with the low inherent Na. However, its impact is less pronounced when the grains are already large in the absence of sodium [30]. Moreover, excessive Na incorporation can be detrimental, as demonstrated by recent studies on CIGS films grown on Mo-coated glass substrates, where moderate NaF co-evaporation led to increased grain size, while very high NaF fluxes resulted in reduced grain size [31]. In contrast, for the thin-ITO absorber T630/(0Na)/ITO100, larger grains are observed despite the absence of intentional Na incorporation. This behavior is attributed to sufficient Na diffusion from the glass substrate during high-temperature growth, facilitated by the reduced ITO thickness and the associated thermal budget. It should be noted that the glass substrate used in this case differs from that in the other

samples and may contain a different intrinsic Na concentration. However, since the exact Na content of the glass was not quantified, its precise contribution cannot be determined and is therefore beyond the scope of the present study.

Figure 2(f) represents the X-ray diffraction patterns of the CIGS absorbers recorded in θ-2θ configuration. The diffraction peaks are indexed to chalcopyrite crystal structure with reference to ICDD (00-056-1309) standard data. A strong peak corresponding to (112) reflection indicates that the absorbers are crystallized into the chalcopyrite phase. Substitution of In with smaller Ga atom results in contraction of the lattice, thereby shifting the (112) peak towards higher angle with increase in [Ga]/([In]+ [Ga]), evident from Figure 2(g). The peak shouldering of (112) orientation in T575/(0Na)/ITO250 absorber is also observed compared to high temperature T630 samples. The chalcopyrite peak in Ga-graded absorbers can be appeared as broad or multiple peaks from different chalcopyrite phases because of elemental intermixing and diffusion [32]. The elemental intermixing depends on various conditions such as, growth process (single-stage or three-stage), growth temperature and presence of Na during the growth. We see that higher growth temperature leads to homogenous phase with a single peak that shifts towards higher diffraction angle which aligns with the existing reports.

All the absorbers were produced by a three-stage process to have Ga grading in the absorber and to control band gap minimum, as described in detail by Damilola et al. [19]. It is well established that the band gap of CIGS absorbers depends on the Ga content of the film [18]. Numerous studies on Se-based chalcopyrite solar cells emphasize the advantages of having more Ga at the back interface, the higher bandgap suppresses back contact recombination [33-36]. For a not perfectly passivated front contact, a V-profile in the bandgap (i.e in the Ga content) is ideal. The bandgap minimum (or "notch") should be not too narrow, to ensure sufficient absorption near the absorption edge [37]. Figure 3(a) presents the GDOES depth profiles of Ga quantified in terms [Ga]/([Ga]+[In] ratio. The GDOES profiles are calibrated using compositions determined from EDX spectroscopy carried out at 20kV.

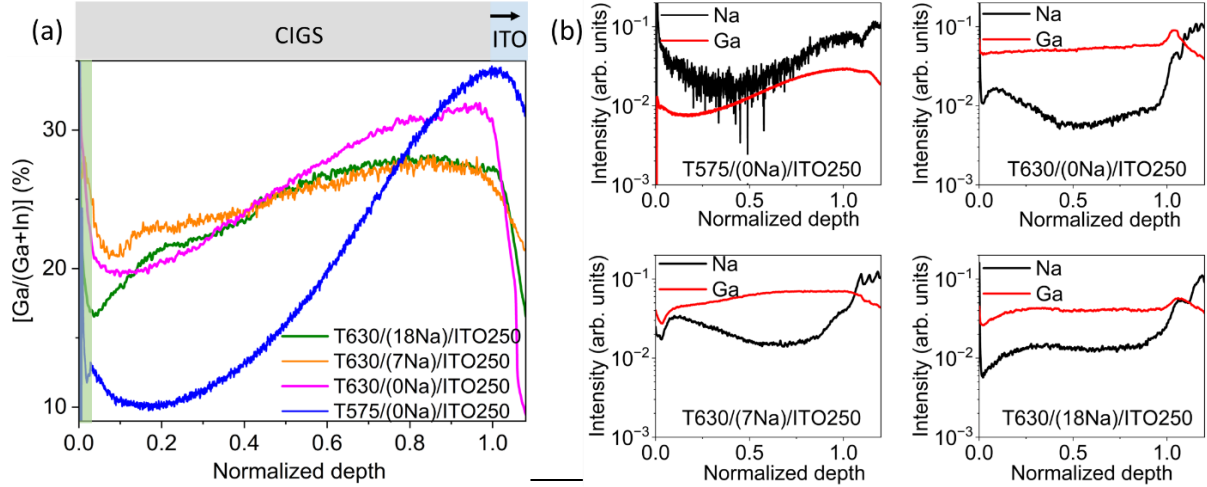

***Figure 3:*** *a) [Ga]/([Ga]+[In]) depth profiles of different CIGS absorbers determined from GDOES. b) Ga and Na signal intensity profiles with normalized depth of the absorbers (real absorber thickness ~ 1.5 μm for all samples). To mention that T575/(0Na)/ITO250 was recorded separately for GDOES than other three. The data for this sample is used only for qualitative comparison to highlight Ga and Na signal changes.*

The GDOES profiles are plotted against normalized depth of the absorber, where 0 indicates surface and 1 refers to the CIGS/ITO interface. The absorber grown at lower temperature has steeper Ga gradient compared to high temperature samples. For sample grown at 575°C, [Ga]/([Ga]+[In]) decreases from 35% at bottom of the film to 10% close to the surface with a broader notch. While the absorbers grown at 630°C have a much flatter gradient and extremely narrow notch because of the higher thermal energy available for interdiffusion of In and Ga during absorber growth. Also, the notch is too close to the surface with little or no increase of Ga content towards the front surface. The Ga flux in the third stage was kept lower than the Ga flux in first stage to have lower Ga at surface. It is shown that high Ga at the surface is associated to low FF, so the processes were designed for lower Ga flux in third stage for sulfide-based chalcopyrite [32]. However, it does not seem to be the case with high temperature absorbers. Here, no significant increase in Ga is seen towards the surface. Therefore, an increase in Ga flux during third stage of the high temperature growth process is expected to result in a better V-shape profile of a notch, which will be explored in next study. Still as we show below, the samples grown at higher temperature outperform the lower temperature samples. These observations indicate that further improvements are possible with an improved Ga gradient for the high temperature growth. Additional Na through co-evaporation is also responsible to facilitate more interdiffusion of In and Ga, as evident by the even flatter Ga gradient for T630/(7Na) and T630/(18Na) absorbers. Na supply from SLG plus additional Na from co-evaporation minimizes lateral band gap fluctuations and enhances In and Ga interdiffusion, studied by Valluvar Oli et. al. [31]. Moreover, a reduction of Ga near the back interface but still inside the absorber is visible, likely due to utilization of Ga for the formation of $GaO_x$ interfacial layer between CIGS and ITO back contact. It is commonly known that processing chalcopyrite absorbers on ITO at high temperatures such as 550°C [8] and even 450°C [5] tends to create a $GaO_x$ interfacial layer. The GDOES intensities of Ga and Na signals are shown in Figure 3(b). The formation of $GaO_x$ at the CIGS/ITO interface is clearly indicated by the abrupt change in Ga signal at the interface. A more pronounced Ga signal is observed for T630/(0Na)/ITO250 compared to T575/(0Na)/ITO250, which can be attributed to the higher growth temperature that enhances Ga diffusion and favors the formation of $GaO_x$. Notably, the Ga signal is much flatter for the absorber with moderate Na incorporation (7Na), while a stronger Ga signal reappears for the absorber with higher Na content (18Na). This difference is particularly notable given that both absorbers were grown under identical substrate temperature and ITO thickness conditions. Although Na incorporation is essential for achieving high CIGS solar cell efficiency, excessive Na supplies, such as that introduced by an overly thick NaF precursor layer has been reported to enhance GaOx formation at the ITO/CIGS interface [8, 12]. This behavior is attributed to the presence of alkali metals during CIGS growth, which promotes Ga segregation toward the interface and facilitates subsequent oxidation.

In parallel, a pronounced Na accumulation is detected at the CIGS/ITO interface. In conventional CIGSe solar cells, such Na accumulation is often associated with Ga-rich regions that promote the formation of smaller grains and a high density of grain boundaries [38]. Alkali elements have also been reported to be preferentially segregated at grain boundaries [39, 40], further contributing to this behavior. In the present case, however, no pronounced layer of fine grains is observed at the back of the absorber. This observation indicates that the high Na signal cannot be solely attributed to grain-boundary segregation. Instead, the accumulation of Na at the interface indicates interaction with, or incorporation into, the GaOx interlayer, which may modify its formation and stability and contribute to the observed non-monotonic dependence on Na supply. Similar observations have previously been reported for CIGSe solar cells with transparent back contacts [8, 41].

GDOES could not be performed on the T630/(0Na)/ITO100 sample due to its unavailability for measurement. However, this sample was included in the time-of-flight secondary ion mass spectrometry (ToF-SIMS) measurements to gain insight into the role of the thin indium tin oxide (ITO) layer in GaOx formation.

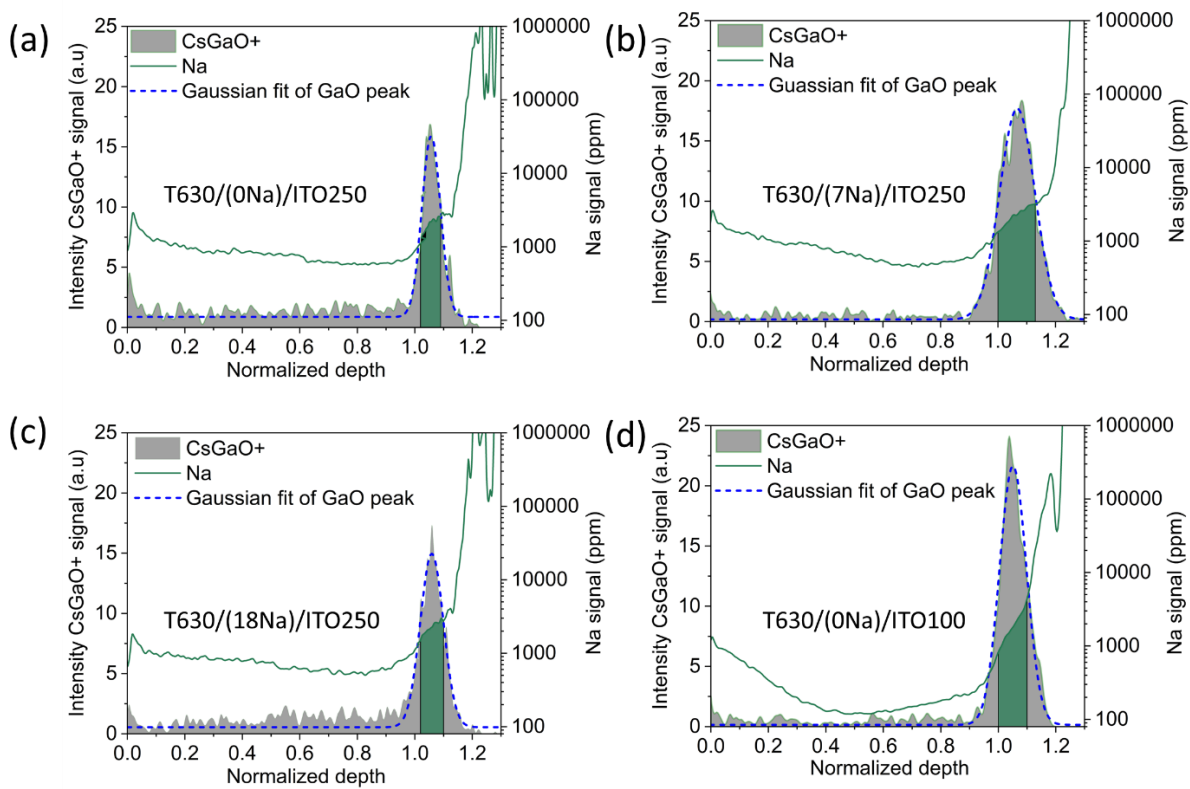

*Figure 4:* ToF-SIMS depth profiles Na and CsGaO signal vs. normalized depth of CIGS absorbers grown at high temperature of 630°C and different Na supply and ITO thickness, (a) T630(0Na)/ITO250, (b) T630(7Na)/ITO250, (c) T630(18Na)/ITO250, and (d) T630(0Na)/ITO100. Green region represents the integral of Na within $GaO_x$ at full width half maximum.

Time-of-flight secondary ion mass spectrometry (ToF-SIMS) was performed on absorbers grown at high substrate temperature (630 °C) to investigate the presence of GaOx and to assess the Na distribution at the CIGS/ITO interface. Figure 4 shows the depth profiles of the $CsGaO^+$ and Na signals as a function of normalized depth. A pronounced $CsGaO^+$ peak is observed at the CIGS/ITO interface for all samples, confirming the formation of a GaOx interlayer and supporting the GDOES observations. Figures 4(a) and 4(d) show a considerable Na concentration in absorbers without intentional Na incorporation, indicating that the ITO layer permits Na diffusion from the glass substrate into the CIGS absorber, evident from GDOES profiles as well. Although the ToF-

SIMS data qualitatively reveals the presence of the GaOx interlayer, the peak width of the $CsGaO^+$ signal cannot be directly used to estimate its thickness due to signal broadening caused by surface and interface roughness. Instead, the GaOx thickness was estimated from the depth difference between the Ga and S signals at 50% of their maximum intensity, as shown in Figure S1. This approach assumes that the depth at which a signal decreases to half its maximum corresponds to the interface between two layers. A delayed decrease of the Ga signal relative to S at the interface indicates the presence of Ga in an oxidized form. The $GaO_x$ thickness was approximated as ~53nm, ~33nm, ~65nm and ~14nm for T630/(0Na)/ITO250, T630/(7Na)/ITO250, T630/(18Na)/ITO250 and T630/(0Na)/ITO100 respectively. For absorbers with thick ITO, high substrate temperature promotes $GaO_x$ formation; however, the $GaO_x$ layer is thinner for the absorber with moderate Na incorporation (7Na) compared to the absorber with higher Na content (18Na). This trend is consistent with the GDOES results, where a more pronounced Ga signal peak is observed at the CIGS/ITO interface for (0Na) and (18Na) absorbers. In contrast, the absorber with thin ITO (T630/(0Na)/ITO100) exhibits the smallest GaOx thickness among the high-temperature samples, which can be attributed to the reduced ITO thickness. In addition to growth temperature, GaOx formation is reported to be strongly influenced by the thickness of the ITO layer as well [9, 25]. The thickness of the ITO layer can influence GaOx formation by acting as a variable oxygen reservoir, with thicker ITO layers providing more oxygen to react with Ga diffusing to the interface at high growth temperatures.

It is to be noticed that Na is lower in the bulk of the absorbers relative to its concentration at CIGS/ITO interface, which further shows steep rise indicating transition of ITO to glass substrate. The thin ITO sample exhibits a significantly lower Na concentration in the bulk of the absorber. Larger grain sizes could partially account for this observation, as Na is known to preferentially accumulate at grain boundaries, consequently, absorbers with larger grains and reduced grain boundary density can exhibit lower Na concentrations in the bulk of the CIGS layer, as previously reported [42]. However, it's noteworthy that the (7Na) and (0Na) samples on thick ITO substrates possess even larger grains which contradicts this attribution. Other possible explanation can be that an increased amount of Na was sucked into the $GaO_x$ layer during its formation, reducing the Na available in the bulk of the absorber. Though, it is not understood yet, why this specifically happened for thin ITO.

A relationship between Na concentration and $GaO_x$ thickness is examined by analyzing the Na distribution within the GaOx interfacial region. To estimate the Na content associated with GaOx, the Na signal was integrated over the full width at half maximum (FWHM) of the Gaussian fit to GaOx related $CsGaO^+$ peak from ToF-SIMS depth profiles. The integrated Na concentrations within the GaOx peak region were estimated to be approximately 185 ppm, 300 ppm, 142 ppm, and 200 ppm for T630/(0Na)/ITO250, T630/(7Na)/ITO250, T630/(18Na)/ITO250, and T630/(0Na)/ITO100, respectively. To facilitate comparison among different absorbers, the Na concentration within the GaOx region was normalized by the estimated GaOx thickness, and the resulting ratio is plotted in Figure 5. The highest normalized Na concentration within GaOx is observed for T630/(0Na)/ITO100, followed by T630/(7Na)/ITO250, T630/(0Na)/ITO250, and T630/(18Na)/ITO250. This trend suggests that Na incorporation within the GaOx interlayer depends not only on the total Na availability but also on the GaOx thickness and interface formation conditions. A higher Na concentration within the GaOx layer may influence its defect density and electronic properties, potentially modifying charge transport across the interface. Previous studies on chalcopyrite absorbers grown on transparent back contacts have reported that Na introduced via external sources, such as NaF post-deposition treatment or NaF precursor layers, can modify the properties of GaOx, with an optimum Na concentration being beneficial [7-9].

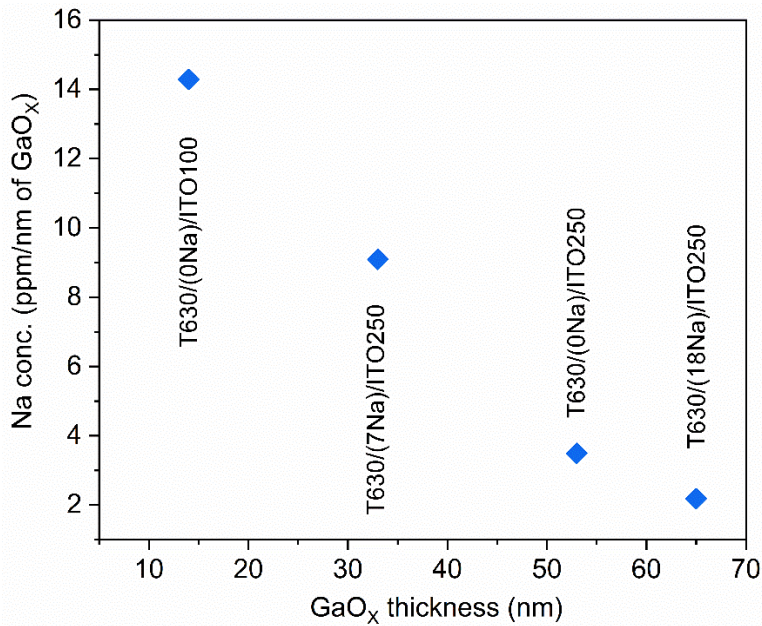

*Figure 5:* Correlation of Na concentration in ppm/nm of $GaO_x$ thickness Vs. $GaO_x$ thickness for different absorbers with or w/o additional Na at high temperature.

### 2.1.2. Optoelectronic properties

Steady state absolute photoluminescence (PL) studies have been performed to evaluate the optoelectronic quality of the absorber. Figure 6(a) shows normalized PL spectra for different absorbers. The PL peak corresponds to band-to-band transition. Thus, as a first approximation, we set the energy of the PL maximum equal to the band gap. Absolute PL flux is also plotted to have better comparison, Figure S2. The observed blue shift in the band gap, from $T_{sub}$ ~575°C to $T_{sub}$ ~630 °C, is attributed to a combination of a weakened Ga gradient and an increased overall Ga content within the notch region (see Fig. 3). This shift results from enhanced elemental interdiffusion at elevated temperatures, leading to a more uniform Ga distribution and consequently a higher band gap [29]. All absorbers grown at high substrate temperature exhibit an increased band gap. However, the relative trend among these samples does not strictly follow the Ga content of the notch inferred from the GDOES measurements (see Fig. 3). This discrepancy is attributed primarily to sample inhomogeneities, which can affect both the local Ga distribution and the optically probed band gap. T575/(0Na)/ITO250 shows a broad emission around 1.1 eV which is attributed to deep defects (D2) and contributes to non-radiative losses. The high temperature sample without additional Na T630/(0Na)/ITO250 shows emission from deep defects D2 and D1 around 1.1eV and 1.4eV respectively, which are well known in sulfide-based chalcopyrite absorbers [21]. Na incorporation in the absorbers T630/(7Na)/ITO250 and T630/(18Na)/ITO250 effectively suppresses D1 and D2 emission, compared to T630/(0Na)/ITO250. The beneficial role of Na in improving CIGS absorber performance has been widely reported and is attributed to mechanisms such as substitution of antisite donor defects, promotion of Cu vacancy formation, and passivation of grain boundary defects [39, 43, 44], all of which contribute to reduced non-radiative recombination. In contrast, the Na-free absorber grown on thin ITO (T630/(0Na)/ITO100) still exhibits signatures of defect-related emissions (D1 and D2). However, a direct comparison of defect emissions between no additional Na absorbers grown on thick and thin ITO layers is not straightforward, as the intensity and nature of these emissions are influenced by multiple factors, including absorber composition, Na availability, and the spatial distribution of constituent elements. Yet, we assume that sufficient Na diffusion from the glass substrate through the thinner ITO layer during high-temperature growth, which enables partial defect passivation even without intentional Na incorporation in (T630/(0Na)/ITO100).

We also deduce QFLS values ($\Delta E_F$) from absolute calibrated PL at photon fluxes equivalent to 1 sun, according to one of the methods described by Siebentritt et al [20]. The method makes use of the relationship between the $\Delta E_F$ of the absorber and $V_{OC}$ in the radiative limit ($V_{OC}^{rad}$), where each photon produces an electron-hole pair that recombines radiatively. We approximate $V_{OC}^{rad}$ by the Shockley-Queisser $V_{OC}$, as tabulated by Ruhle et. al [45], that would be obtained with a band gap equal to PL maximum. This assumption overestimates the radiative loss, i.e. underestimates $V_{OC}^{rad}$ and $\Delta E_F$ by about 10 to 20meV [46]. $\Delta E_F$ is then determined from equation (1),

$$\Delta E_F = qV_{OC}^{rad} + k_B T ln(Y_{PL}) \qquad (1)$$

where $k_B T ln(Y_{PL})$ describes the non-radiative voltage loss $\Delta V_{OC}^{nr}$ with $k_B$ the Boltzmann constant and $T$ the temperature. $Y_{PL}$ is the photoluminescence quantum yield which can be calculated as $Y_{PL} = \emptyset_{PL}/\emptyset_{inc.}$, where $\emptyset_{PL}$ is the integrated photon flux of the band-to-band PL emission peak and $\emptyset_{inc.}$ is the incident photon flux of the laser. Because reflection losses are not considered, this approach underestimates $Y_{PL}$ somewhat and leads to a further underestimation of $\Delta E_F$ by about 5 meV. Thus, the $\Delta E_F$ values discussed here are a bit lower than the actual QFLS under 1 sun illumination.

Figure 6(b) compares different parameters quantified from PL analysis of all absorbers and corresponding values are tabulated in Table S2 in supplementary information. $\Delta E_F$ of the absorbers significantly improved for $T_{sub}$ ~630 °C samples with a large gain of 180 meV in T630/(0Na)/ITO250, resulting in QFLS of 1060 meV compared to 880 meV for T575/(0Na)/ITO250. $Y_{PL}$ is also enhanced by two orders of magnitude, which further improves with additional Na, in case of T630/(7Na)/ITO250 and T630/(18Na)/ITO250. This improvement can be attributed to reduction in non-radiative losses. Although the relative suppression of defect emissions cannot be directly compared between thin- and thick-ITO absorbers, their similar $Y_{PL}$ suggests comparable non-radiative losses.

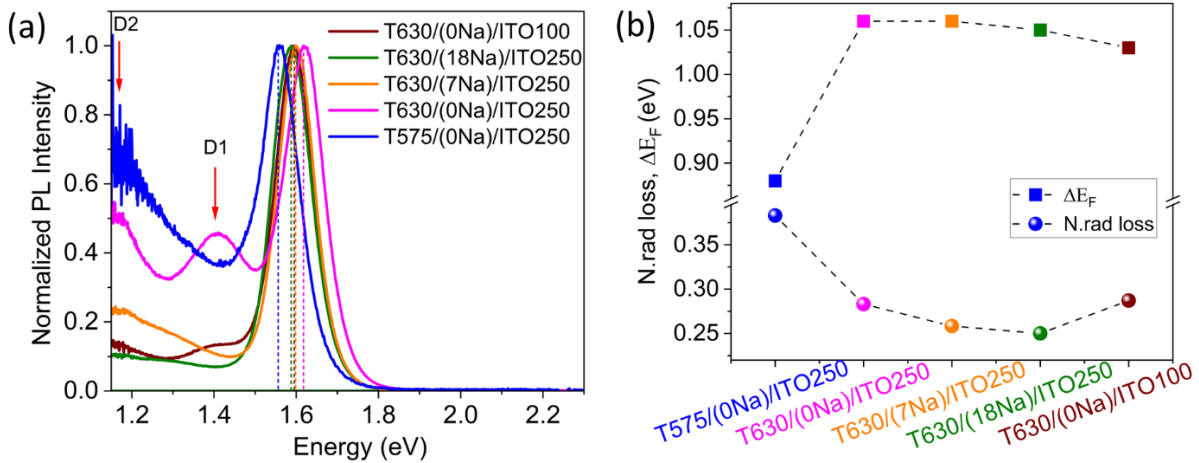

*Figure 6:* a) Normalized PL emission spectra of the absorbers, where vertical dashed lines mark the PL maximum, b) comparison of $\Delta E_F$ and non-radiative voltage loss among different absorbers.

## 2.2 Solar cell performance

Solar cells were fabricated with solution grown CdS buffer layer followed by sputtering the Al:ZnMgO/Al:ZnO window layer then followed by Ni/Al metal grids for the top contact. The current density vs. voltage (J-V) characteristics of the ITO/CIGS solar cells are displayed in Figures 7(a) and 7(b). T575/(0Na)/ITO250 and T630/(0Na)/ITO100 show single diode characteristics, while the J-V curve of T630/(0Na)/ITO250 exhibits an S-shape (double-diode), Figure 7(a). Comparing the samples at high temperature and without additional Na shows that the device on thinner ITO demonstrates better performance with no distortion in the J-V curve. The J–V curve of T630/(0Na)/ITO250 exhibits a pronounced S-shaped distortion, which is typically associated with a carrier extraction barrier at the back contact. At low forward bias, the device follows the expected diode behavior governed by the primary absorber/buffer junction. However, at higher forward voltages, the current deviates from ideal diode behavior, indicating the presence of an additional barrier that limits carrier extraction. Such behavior is commonly attributed to a back-contact extraction barrier, in this case at the CIGS/ITO interface. For low temperature sample, no such shape is observed. The two devices with additional Na T630/(7Na)/ITO250 and T630/(18Na)/ITO250 compared to without additional Na absorber T630/(0Na)/ITO250 separately, in Figure 7(b). The sample with higher Na supply leads to an S-shape J-V characteristic i.e low FF similar to without additional Na device. Both the samples indicate an additional barrier hindering the functioning of the primary diode CIGS/CdS junction, which could be correlated to the formation of GaO$_x$. Although GaOx was detected at the CIGS/ITO interface in all absorbers, the barrier appears significantly stronger in certain devices, where it manifests as a kink in the J–V characteristics. To understand this behavior, we correlate the electrical response with depth profiles obtained from GDOES and ToF-SIMS. Notably, devices exhibiting a kink show a substantially thicker GaOx layer, indicating the formation of an extraction barrier at the back contact. This thicker oxide likely impedes carrier transport across the CIGS/ITO interface, thereby limiting efficient charge extraction. In contrast, devices without a kink exhibit a thinner GaOx layer, resulting in less impeded carrier extraction. Similar observations have been reported previously, where the GaOx thickness was found to strongly influence the J–V characteristics of CIGS devices [25]. As the GaOx thickness was not directly quantified for the low-temperature devices, it is assumed to be too thin to significantly influence charge transport, owing to the reduced thermal budget, as outlined in the Introduction.

The statistical comparison of six cells for each growth condition is represented in supplementary information Figure S3. The improvement in the $V_{OC}$ open circuit voltage for high temperature devices is mainly due to the improved $\Delta E_F$ as observed from Figure 7(c), except for T630/(18Na)/ITO250 where a strong transport barrier at the back contact reduces overall performance. Although T630/(18Na)/ITO250 exhibits the thickest GaOx interlayer at the back interface, it shows the best absorber properties among others, with minimal non-radiative losses and the highest $Y_{PL}$. For absorbers with short carrier diffusion lengths such as CIGS, PL primarily probes intrinsic bulk recombination and is minimally influenced by the back interface. Consistent with this, previous studies have reported comparable PL intensities in CIGS measured with and without back contact under front-side excitation, suggesting minimal influence of the back interface on PL emission [47]. A similar behavior is expected here, and therefore the PL emission is unlikely to be significantly affected by the GaOx layer at the back interface. However, the back interface remains critical for device performance, as charge carriers reaching the interface must overcome the associated barrier, which can limit carrier collection and overall device efficiency.

The interface loss, quantified as the difference between $\Delta E_F$ and $V_{OC}$, is due to a gradient in the minority carrier quasi-Fermi level caused by interface recombination, which reduces $V_{OC}$ more strongly than $\Delta E_F$ [48]. Interface loss is found to be relatively lower for T575/(0Na)/ITO250 and T630/(0Na)/ITO100, however the performance is mainly limited due to the non-radiative losses in the bulk for all the absorbers. It is important to mention that $V_{OC}$ in all devices is also limited by

CdS buffer layer, most probably due to non-ideal band alignment with wide-gap CIGS absorber. The current density has been reduced for cells deposited at high temperature, this loss could be partially due to band gap widening, as can be seen from the onset of EQE in Figure 7(d). The low temperature device (blue curve) extended to the longer wavelengths, which improves the overall carrier collection, see Figure S4. Also, relatively wider notch in low temperature absorber could be responsible for better absorption. The FF is the lowest for the cells grown at high temperature without additional Na T630/(0Na)/ITO250 and with additional Na T630/(18Na)/ITO250, which is due to the observed S-shape of their J-V characteristics. Device with moderate Na i.e T630/(7Na)/ITO250 and device on thin ITO without additional Na i.e T630/(0Na)/ITO100 have shown the highest PCE among all devices with FF > 70%.

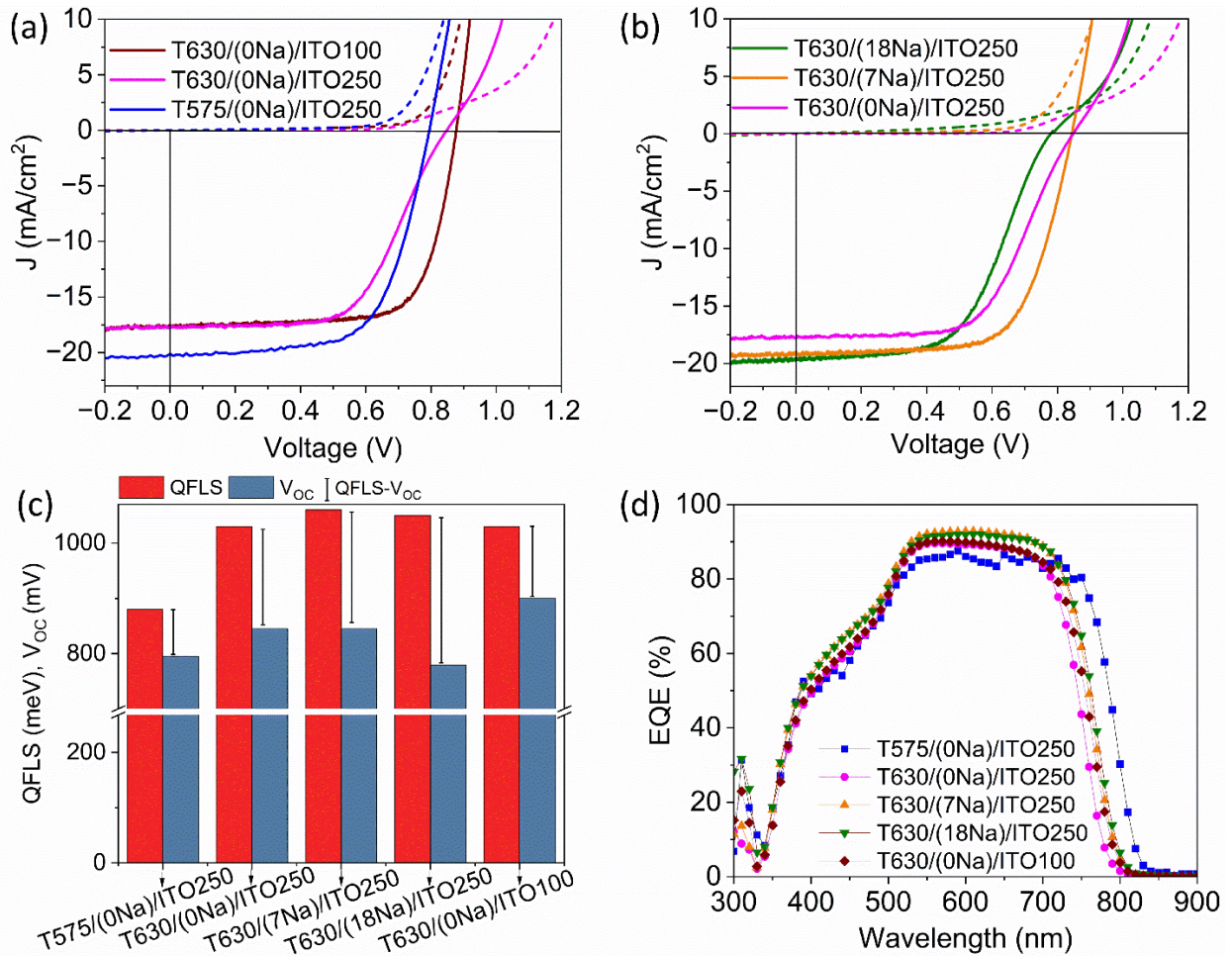

*Figure 7:* J-V characteristics of the best cell on each sample in the dark and under light (w/o ARC), (a) comparison between T575, T630 without NaF on ITO(250) and T630 without NaF on

ITO(100), (b) comparison of T630(0Na), T630(7Na) and T630(18Na), (c) QFLS and $V_{OC}$ comparison, (d) EQE spectra of different devices.

It may be due to the thin layer of $GaO_x$ and Na presence at the interfacial layer which makes some shunting paths to facilitate the charge transport. It indicates that presence of $GaO_x$ at CIGS/ITO interface is not always detrimental, however, the properties of $GaO_x$ layer are crucial in deciding the performance of CIGS solar cells.

*Table 2. Solar cell performance of the wide-gap CIGS champion cells with ARC. One with thin ITO without any additional Na source and other with thick ITO with 7 min NaF co-evaporation.*

| Absorber/device | $E_g^{EQE}$ (eV) | $J_{SC}^{EQE}$ ($mA/cm^2$) | $V_{OC}$ (mV) | FF (%) | Active area PCE (%) |
|---|---|---|---|---|---|
| T630/(7Na)/ITO250 | 1.61 | 20.4 | 865 | 68.5 | 12.0 |
| T630/(0Na)/ITO100 | 1.61 | 19.7 | 900 | 71.8 | 12.7 |

The two best performing devices were equipped with an $MgF_2$ anti-reflective coating (ARC). Their PV parameters are presented in Table 2 and Fig. 8. The CIGS band gaps are calculated from the peak position of dEQE/dE vs. E plots. J-V characteristics with ARC for these devices are shown in Figure 8(a). One of the champion devices is also tested under rear illumination conditions. The efficiency under rear illumination is significantly lower (3.2%) than the front (12.7%), as shown in Figure S5. The efficiency from the backside is mainly limited due to low diffusion lengths and reflection losses at CIGS/ITO interface. The device on thin ITO without additional NaF is observed to have lowest interface loss (140 meV) better band match buffer/CIGS interface, which is subject to verification. However, it exhibits a slightly higher non-radiative loss than T630/(7Na)/ITO250. Finally, we compare our best results on ITO with other CIGS technologies with transparent back contacts, as shown in Figure 8(b). This work is the first report to demonstrate this level of efficiency on wide band gap (>1.5 eV) chalcopyrite material with transparent back contact. Wide-gap devices lose current by huge factor than the gain in $V_{OC}$, therefore efficiency shows a decreasing trend with increasing band gap [49]. However, our results with band gap 1.6 eV are comparable to selenide devices around 1.45 eV. Wide gap sulfide chalcopyrite at temperature growth process helped us to achieve decent absorber properties with band gap of 1.6eV, which leads to nearly 13% PCE. This requirement poses challenges, as elevated temperatures with transparent back contact can exacerbate interfacial issues, including altering the properties of the back contact and GaO$_x$ formation. The role of Na on modifying GaO$_x$ properties is still not fully understood. Further investigation of Na incorporation in thin-ITO absorbers, through controlled variation of Na concentration, would be valuable.

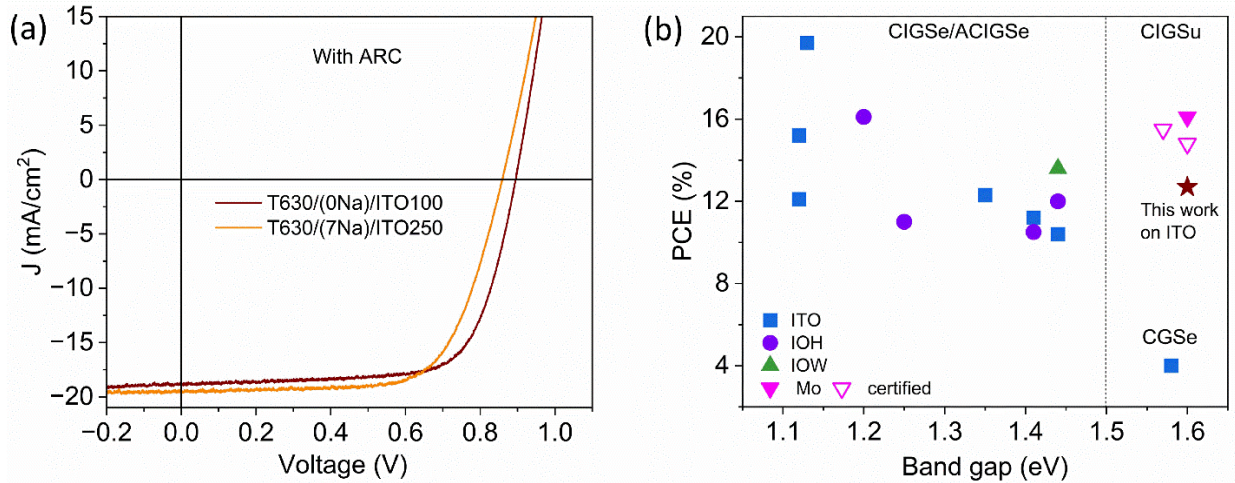

*Figure 8: (a) J-V characteristics of the wide-gap CIGS champion cells after ARC, (b) Efficiency vs. band gap comparison of chalcopyrite solar cells on different TBC in literature[3, 5, 6, 8, 10-12, 19], with our best result. The devices on Mo indicate the record efficiencies of sulfide chalcopyrite with non-transparent back contact.*

3. Conclusion

This study demonstrates that Na incorporation and ITO thickness strongly influence GaOx interlayer formation, defect passivation, and optoelectronic properties of wide-bandgap CIGS absorbers on transparent back contacts. High temperature CIGS growth led to improved optoelectronic properties and QFLS above 1.0 eV at a band gap of 1.6 eV. An additional Na implementation via NaF co-evaporation, beside the Na provided from the SLG substrate, was found to reduce the non-radiative losses by suppressing deep defects. Even in the absence of intentional Na supply, absorbers grown on thin ITO exhibit comparable optoelectronic quality, attributed to Na diffusion from the glass substrate during high-temperature growth. These observations confirm the critical role of Na in improving bulk absorber quality through defect passivation. Depth profiling from GDOES and ToF-SIMS confirms the presence of a GaOx interlayer, with its thickness showing a non-monotonic dependence on Na concentration. While clear correlations between Na supply, GaOx formation, and optoelectronic properties are established, the exact mechanisms by which Na modifies interfacial oxidation and defect chemistry are not yet fully understood. In particular, the interplay between Na accumulation, Ga redistribution, and oxygen diffusion at the CIGS/ITO interface requires further investigation.

The PCE of 12.7% with a high FF of 71.8% for a band gap of 1.6 eV is realized for CIGS/ITO solar cell, which is the highest reported PCE for chalcopyrite with $E_g > 1.5$ eV on TBC, to date. This paper demonstrates the potential of wide gap chalcopyrite thin-film solar cells based on sulfides as top cell in tandem solar cell designs.

**Experimental details**

The CIGS absorbers were grown by co-evaporation on SLG/ITO substrates. Two different ITOs have been used in this study, in-house deposited 250 nm ITO and 100 nm commercial ITO from SOLEMS. High-temperature resistant SLG (from SAIDA) glass has been used to carry out high temperature growth. A three-stage growth process described in Ref [19] is used to produce high quality absorbers. The actual substrate temperature in first and second/third stage is 475°C and 630°C, respectively. The thermocouple for substrate temperature is located at the back of the heater which gives a discrepancy between the actual temperature of the sample surface from pyrometer readout and the set temperature substrate. The substrate temperature was calibrated by correlating the set temperature and to the actual temperature of the substrate from pyrometer readout, which was calibrated to the softening point of a glass substrate with known softening temperature. The sample stage was continuously rotated at 8 rpm to have homogenous films. The schematic of the growth process is represented in Figure 9.

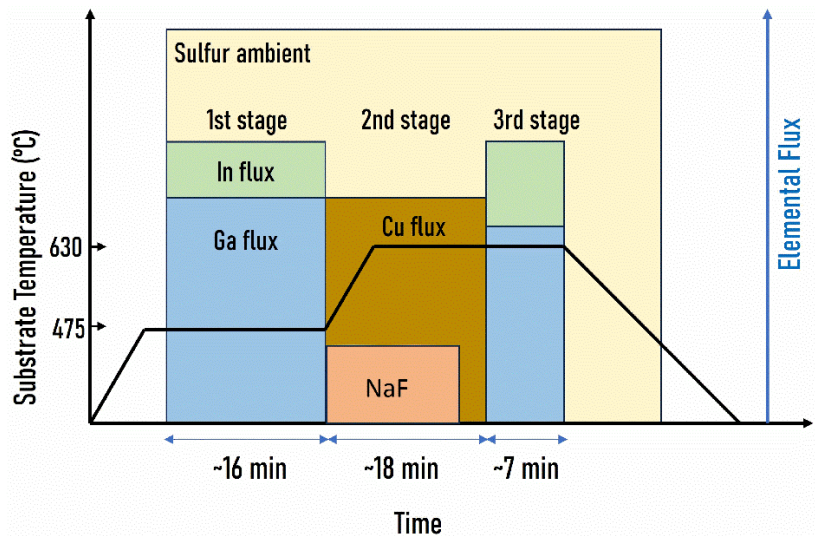

*Figure 9: The deposition profile of the CIGS absorbers grown with three-stage process. Axes (black) indicate temperature and (blue) elemental fluxes. NaF co-evaporation in the second stage to have additional Na in the samples. Ga flux in the third stage is reduced compared to the first stage.*

The first stage includes the evaporation of In and Ga under sulfur ambient with an average pressure of $6\times10^{-3}$ mbar. The second stage involves the Cu and NaF evaporation to have additional Na in some CIGS absorbers at slightly higher sulfur pressure. The deposition time for NaF varied to have different Na content. Metal source fluxes were calibrated using electron impact emission spectroscopy (EIES). Samples were allowed to cool naturally within sulfur ambient after the deposition process ends. The time duration of the deposition is optimized to get 1.5 µm thick absorbers. After absorber deposition, 60 nm thick CdS buffer layer was deposited by conventional chemical bath deposition. No chemical etching was performed on the absorber before buffer layer deposition. The solar cells were finished by depositing an Al:MgZnO (80 nm) window layer, followed by an Al:ZnO (380 nm) transparent front contact, both applied using magnetron sputtering techniques. Subsequently, a Ni/Al grid electrode was thermally evaporated onto the structure. A 100 nm thick $MgF_2$ antireflection coating was deposited on the best devices. Individual cells were delineated through mechanical scribing, resulting in an active area of approximately 0.45 cm².

**Characterization details**

GDOES was used to determine the compositional depth profiles of the CIGS absorbers. The sputtered atoms' light emission is produced by eroding the sample in an argon plasma at 450 Pa of pressure. Several photomultipliers detect the optically diffracted emission lines of the single elements. The reference measurement using EDX was used to quantify the CIGS matrix elements. A relative sensibility factor (RSF), which was computed using implanted reference samples, is used to determine the sodium content.

ToF-SIMS measurements were carried out with a ToF5-SIMS instrument from IONTOF. The analyzing $Bi^+$-ion beam was run at 30 keV and probed over an area of $50 \times 50$ µm². Furthermore, a Cs ion gun with 2 keV and a rasterized area of $200 \times 200$ or $250 \times 250$ µm² was used to reach good sputter rates and to detect Cs-clusters as e.g. CsGaO, which are nearly independent to matrix effect.

The quasi-fermi level splitting measurements on the bare absorbers were performed by a home-built absolute PL set-up. The system employed a 405 nm laser as the excitation source and operated at room temperature. Emission spectra were captured using a silicon charge-coupled device (CCD) detector. Prior to the measurements, the experimental setup is first calibrated to ensure both spectral and intensity accuracy. The calibration procedure follows as described in ref. [47]. The incident photon flux was meticulously adjusted to replicate the standard AM1.5 solar spectrum, corresponding to the photon flux that an absorber with a bandgap of 1.6 eV would experience under one sun illumination. The PL quantum yield method was used to determine the $\Delta E_F$ using the PL maximum as the band gap and explained in section 2.1.2.

The photovoltaic performance of the solar cells was evaluated under standard testing conditions using a class AAA solar simulator, calibrated with a certified silicon reference cell (RQN3154). I–V characteristics were recorded at room temperature in the forward scan direction using a source-measure unit, with a scan rate of 50 mV s$^{-1}$. The active cell area, estimated to be ~0.45 cm², was determined using a Leica optical microscope and quantified via *ImageJ* software. This area was defined by the outer boundary of the mechanically scribed cell and included the grid-shaded regions, as no masking was applied during illumination. The entire device area was exposed to AM1.5G equivalent illumination at 100 mW cm$^{-2}$ (1 Sun). External quantum efficiency (EQE) measurements were performed using modulated light from a halogen–xenon lamp (Ushio UXL-302-0), with the resulting photocurrent detected by a lock-in amplifier.

**Acknowledgement**

This work is a part of the SITA project (101075626), funded by the European Union. Views and opinions expressed are however those of the authors only and do not necessarily reflect those of the European Union or CINEA. Neither the European Union nor the granting authority can be held responsible for them. Part of this work was funded by the Luxembourgish Fond National de la Recherche FNR in the framework of the REACH project (INTER/UKRI/20/15050982). For the purpose of open access, the author has applied a Creative Commons Attribution 4.0 International (CC BY 4.0) license to any Author Accepted manuscript version arising from this submission.

# Conflict of interest

There are no conflicts of interest to declare.

# Supplementary Information

# Near 13% efficient semitransparent Cu(In,Ga)S$_2$ solar cells with band gap of 1.6 eV on transparent back contact

Kulwinder Kaur[1]*, Arivazhagan Valluvar Oli[1], Michele Melchiorre[1], Wolfram Hempel[2], Wolfram Witte[2], Jan Keller[3], Susanne Siebentritt[1]

[1]Laboratory for Photovoltaics, Department of Physics and Materials Science, University of Luxembourg, Luxembourg

[2]Zentrum für Sonnenenergie- und Wasserstoff-Forschung Baden-Württemberg (ZSW), Germany

[3]Ångström Solar Center, Division of Solar Cell Technology, Uppsala University, Sweden

*Corresponding author: kulwinder.kaur@uni.lu

**Table S1:** Chemical composition measured in bulk (20 kV) and near-surface region (7 kV) using EDS for all the CIGS absorbers.

| Absorber | [Ga]/([Ga]+[In]) | | [Cu]/([Ga]+[In]) | |
|---|---|---|---|---|
| | 20 kV | 7 kV | 20 kV | 7 kV |
| T575/(0Na)/ITO250 | 0.20 | 0.13 | 1.14 | 0.99 |
| T630/(0Na)/ITO250 | 0.25 | 0.27 | 0.87 | 1.00 |
| T630/(7Na)/ITO250 | 0.23 | 0.22 | 0.87 | 0.95 |
| T630/(18Na)/ITO250 | 0.24 | 0.21 | 0.89 | 0.96 |
| T630/(0Na)/ITO100 | 0.24 | 0.22 | 0.89 | 1.03 |

**Table S2**: CIGS absorbers with different growth temperature and Na supply with their corresponding band gap, QFLS, non- radiative V$_{OC}$ loss, and photoluminescence quantum yield.

| Sample name | ITO Thickness | $E_g^{PL}$ | QFLS | $\Delta V_{OC}^{nr}$ | $Y_{PL}$ |
|---|---|---|---|---|---|
| T575/(0Na)/ITO250 | 250 nm | 1.53 eV | 880 meV | 383 meV | $3.0\times10^{-7}$ |
| T630/(0Na)/ ITO250 | 250 nm | 1.62 eV | 1060 meV | 283 meV | $1.2\times10^{-5}$ |
| T630/(7Na)/ ITO250 | 250 nm | 1.60 eV | 1060 meV | 258 meV | $3.2\times10^{-5}$ |
| T630/(18Na)/ ITO250 | 250 nm | 1.58 eV | 1050 meV | 250 meV | $4.5\times10^{-5}$ |
| T630/(0Na)/ ITO100 | 100 nm | 1.59 eV | 1030 meV | 287 meV | $1.1\times10^{-5}$ |

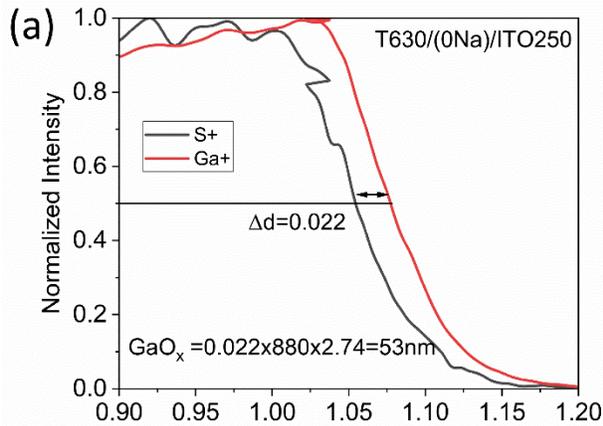
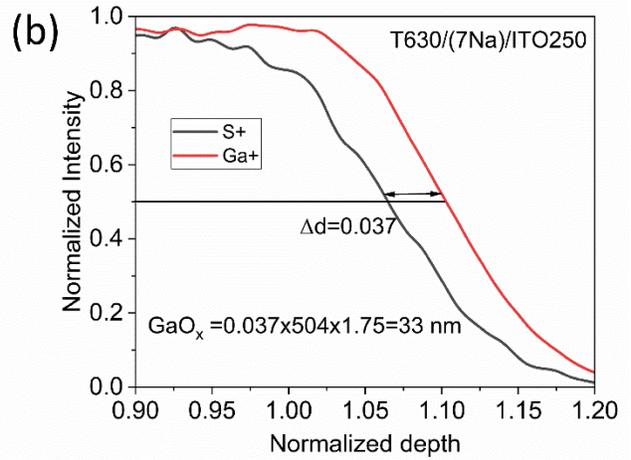
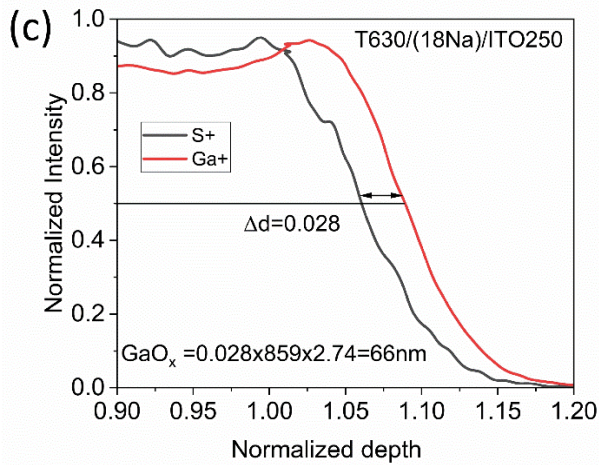
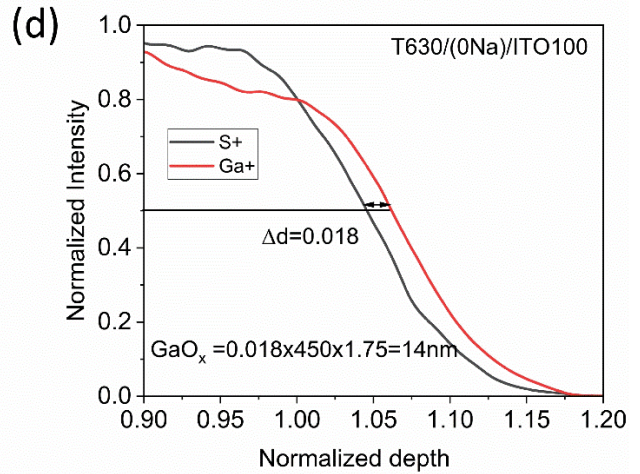

**Figure S1.** Estimation of GaO$_x$ thickness from the time difference between the Ga and S signals at 0.5 normalized intensity.

**GaO$_x$ thickness calculation:**

Thickness = Δd × T × sputter rate. Where Δd is the normalized depth difference at 0.5 normalized intensity, T is the time corresponding to normalized depth of 1. The sputter rate is different for two samples than other two because of different crater size on different day of measurement.

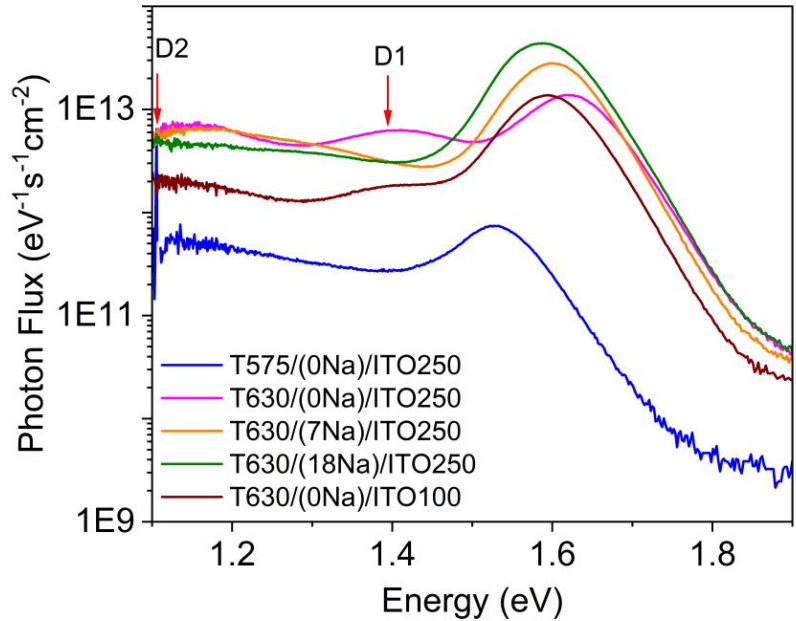

**Figure S2:** Absolute PL photon flux of different absorbers indicating enhanced PL emission and reduced defect emission with additional Na at high temperature.

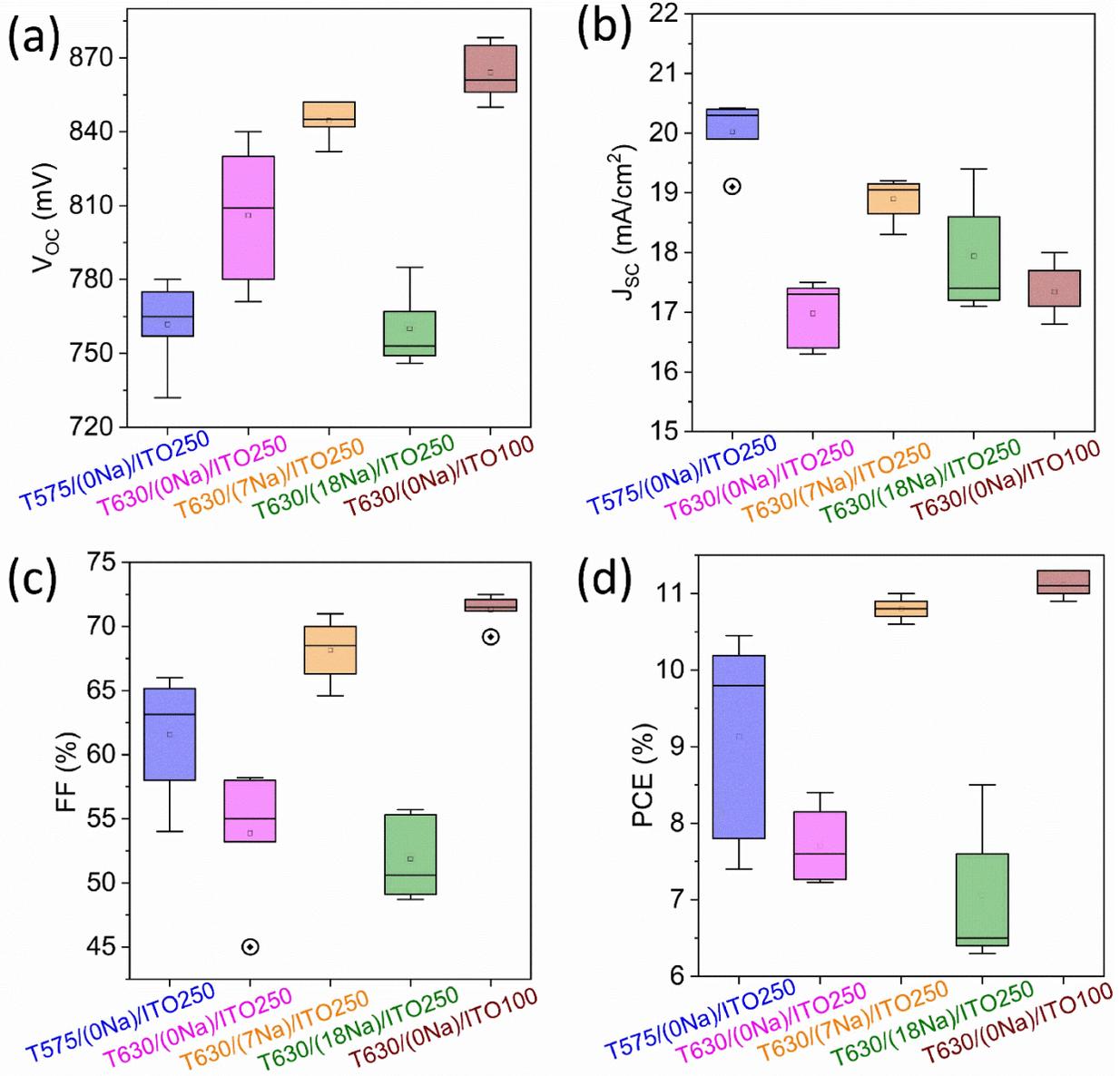

**Figure S3** Statistical comparison of CIGS solar cell parameters of five different cells from the different samples, (a) $V_{OC}$, (b) $J_{SC}$, (c) FF, and (d) PCE.

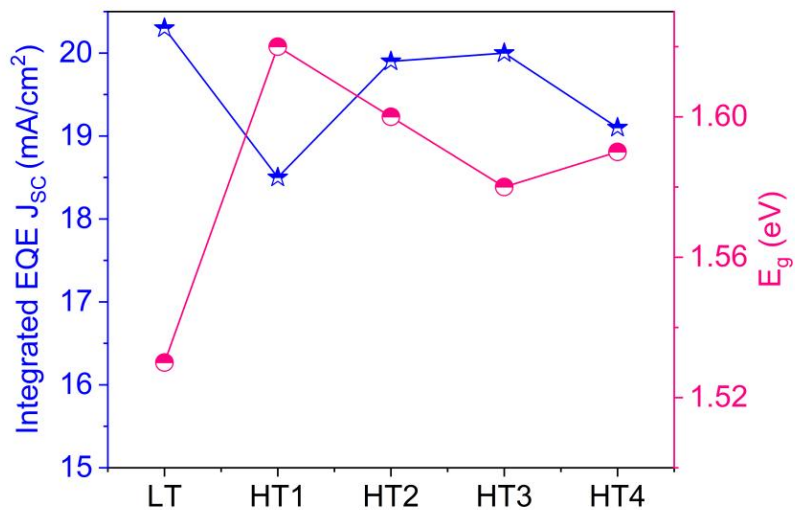

**Figure S4:** Integrated $J_{SC}$ and band gap are plotted for different absorbers grown at low and high temperature. The absorber LT, HT1, HT2, HT3, and HT4 corresponds to T575/(0Na)/ITO250, T630/(0Na)/ITO250, T630/(7Na)/ITO250, T630/(18Na)/ITO250 and T630/(0Na)/ITO100

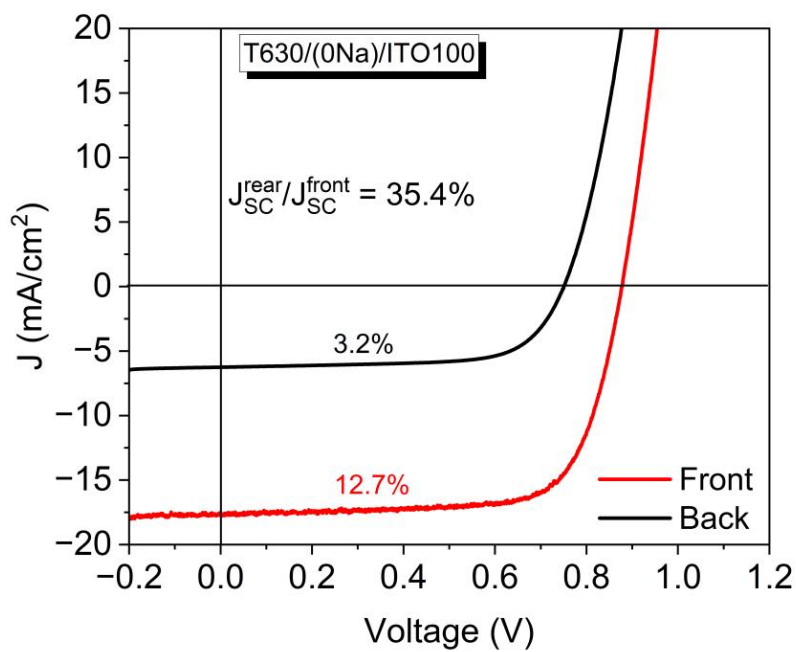

**Figure S5:** J-V curve of the champion cell of T630/(0Na)/ITO100 under front and back illumination condition.